\documentstyle[graphicx,times,fleqn]{jaa}

\newcommand{\disp}{\displaystyle}

\begin{document}

\title[Orbital signatures from blazars]{Orbital signatures from observed light curves of blazars}

\author[A.\ Mangalam \& P.\ Mohan]{A.\ Mangalam$^{1}$ and P. \ Mohan$^{1}$  \thanks {E-mail: (mangalam, prashanth)@iiap.res.in}\\ $^{1}$Indian Institute of Astrophysics, Sarjapur Road, Koramangala, Bangalore, 560034, India}
\date{Accepted . Received ; in original form }
\pagerange{\pageref{firstpage}--\pageref{lastpage}} \pubyear{2013}
\maketitle
\label{firstpage}

\begin{abstract}
Variability in active galactic nuclei is observed in ultraviolet to X-ray emission based light curves. This could be attributed to orbital signatures of the plasma that constitutes the accretion flow on the putative disk or in the developing jet close to the inner region of the central black hole. We discuss some theoretical models which build on this view. These models include general relativistic effects such as light bending, aberration effects, gravitational and Doppler redshifts. The novel aspects relate to the treatment of helical flow in cylindrical and conical geometries in the vicinity of a Schwarzschild black hole that leads to amplitude and frequency modulations of simulated light curves as well as the inclusion of beaming effects in these idealized geometries. We then present a suite of time series analysis techniques applicable to data with varied properties which can extract detailed information from them for their use in theoretical models.
\end{abstract}

\section{Introduction}
Short-timescale variability (1000 - 10000 s) in optical/ultraviolet (UV) and X-ray light curves (of typical duration $\sim$ 25000 s) of active galactic nuclei (AGN) could originate from an accretion disk or jet. Soft \& hard X-rays are believed to be optical/UV radiation reprocessed by a hot, optically thin corona surrounding the inner disk.
Possible extents of this inner region and a break frequency as an orbital signature are presented in \S 2.
We briefly review existing models in \S 3 and present our preliminary results from disk-jet models. Time series analysis of observed light curves (LCs) can detect quasi-periodic oscillations (QPOs), provide constraints on mass and spin of the supermassive black hole (SMBH) and infer the physics of the emission region. Related numerical experiments, a periodicity search strategy, its application and the verification of the LSP and wavelet techniques are presented in \S 4.  

\section{Constraints on emission region for Kerr BH and break frequency}
\begin{figure}
\includegraphics[scale=.22]{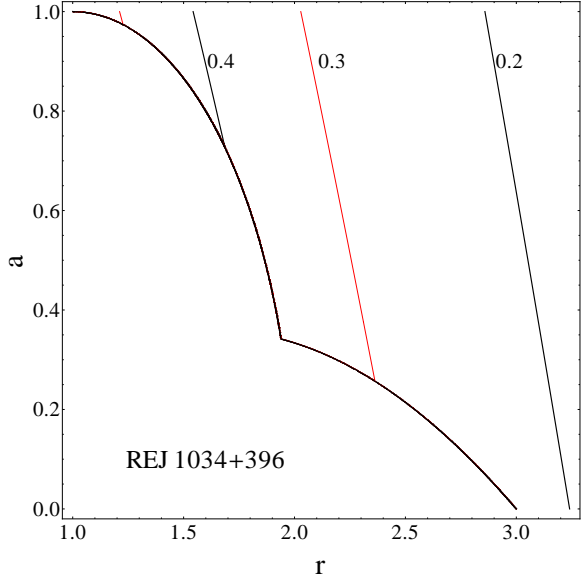}
\caption{\small The allowed region in the parameter space of $r$ and $a$ (in units of $M$) for stationary observers outside the horizon radius lies to the right of the constructed curves representing the equality limits of the inequalities (\ref{eqn1} \& \ref{eqn2}). The SMBH mass contours (in units of $10^8 M_\odot$) are plotted in this allowed region for REJ 1034+396 ($z$=0.043) using $\displaystyle{\frac{M_{\bullet}}{M_\odot}=\frac{3.23 \times 10^4 P}{[r^{\frac{3}{2}}+a](1+z)}}$ with $P$=3767 s (QPO).}
\label{fig1}
\end{figure}

Emission could at the most arise from sources near the innermost stable circular orbit (ISCO) for a Keplerian disk around the SMBH.
The constant spin $\Omega$ in a rotating metric must obey $\Omega_{\rm min} < \Omega <\Omega_{\rm max}$ where 
\begin{equation}
\Omega_{\rm max, min} ={-g_{t \phi} \pm (g^2_{t \phi} -g_{tt} g_{\phi \phi})^{\frac{1}{2}} \over g_{\phi \phi}}.
\end{equation}
For the Kerr case this leads to the condition that,
\begin{eqnarray}
4 a^4 r+8 a^4+2 a^3 r^{\frac{7}{2}}+8 a^3 r^{\frac{5}{2}} +8 a^3 r^{\frac{3}{2}}+a^2 r^5 \nonumber \\
-a^2 r^4 -2 a^2 r^3 +2 a r^{\frac{11}{2}}+4 a r^{\frac{9}{2}}+r^7-3 r^6 > 0,
\label{eqn1}
\end{eqnarray}
where $r$ and $a$ are in units of $M$. In addition, we have the condition that the emitting region lie outside the event horizon of the hole given by,
\begin{equation}
r > r_+=1+\sqrt{(1-a^2)}.
\label{eqn2}
\end{equation}
Conditions (\ref{eqn1}) and (\ref{eqn2}) can be used together to place lower limits on the emission radius and upper limits on the black hole mass assuming that emission is from an orbital feature (see Fig \ref{fig1}). Imprints of the ISCO would be left on outward bound emission. A high frequency break in the periodogram observed in black hole X-ray binaries (eg. Papadakis et al. 2002) could be a possible signature. A break frequency is inferred from the broken power law model, the choice of which among many is dictated by statistical analysis (eg. Mohan et al. 2013 in this volume). The break timescale $T_B$ is related to the location of $r_{ISCO}$ and is given by:
\begin{eqnarray}
T_B = \frac{1}{\Omega_8} M_8 f(a) \ {\rm days} = 0.86 M_8 f(a) \ {\rm hours} \\ \nonumber
f(a) = (r^{\frac{3}{2}}_{ISCO}(a)+a);\\ \nonumber
\Omega_8 = \displaystyle \frac{86400 \ c^3}{2 \pi G M_{\odot} \times 10^8} = 27.832; \\ \nonumber
\end{eqnarray}
for $r_{ISCO}$ as given in eg. Shapiro \& Teukolsky (1983). The spin $a$ can be related to the accretion history of the SMBH and is obtainable from a self consistent model involving $M_8$, $a$, $\dot{m}$, $L_{jet}$, $L$ for an evolving AGN (in preparation). Using these, $T_B$ can be calculated and compared with observations.

\section{Disk-jet variability variability models}
Early models of optical/UV and X-ray variability are discussed in Wiita et al. (1991), Zhang \& Bao (1991), Mangalam \& Wiita (1993) and references therein. Imprints on the emission due to time delay, gravitational and Doppler shift, eclipsing and disk structure and intrinsic variability are cast in terms of an effective redshift factor $\disp{g = \frac{E_{\mathrm{obs}}}{E_{\mathrm{em}}}}$, the ratio of observed to emitted energy of the photon and an integrated flux $\disp{F=I_{\mathrm{em}} g^4 d\Omega}$ where $I_{\mathrm{em}}$ is the bolometric emission intensity and $d\Omega$ is the solid angle of the source subtended at the observer. These models provide a reasonable match to observations: a high frequency power law region, a low frequency flattening possibly due to decrease in coherency of orbital features and a high frequency cut-off due to the presence of the disk inner edge (assumed to be at the ISCO) beyond which emission ceases as material enters the black hole. Results from simulations carried out by Mangalam \& Wiita (1993) are in agreement with optical/UV based power spectra and predict power laws with slopes in the range -1.1 to -2.4. In recent models, light bending (eg. Beloborodov 2002, Pechacek et al. 2005) and the corresponding time delay (Pech{\'a}{\v c}ek et al. 2006) are considered. Jet based models relevant to the current study include the Doppler beaming of emitting sources on helical paths (eg. Camenzind \& Krockenberger 1992, Steffen et al. 1995, Rieger 2004). The inner jet appears to be intrinsically linked to the corona as its spectral characteristics are similar to that from a Compontized corona for X-ray binaries (Markoff et al. 2005). The VLBI spectrum of the inner jet of M87 indicates a compact emitting region, even inside of the ISCO, implying that its source is on the disk (Doeleman et al. 2012). As perturbations produced on the disk can be advected into jets and amplified there by Doppler boosting (e.g. Wiita 2006), it is important to study the inner disk and developing jet with foot points on the disk.

\begin{figure}
\centerline{\includegraphics[height=6cm,width=7cm]{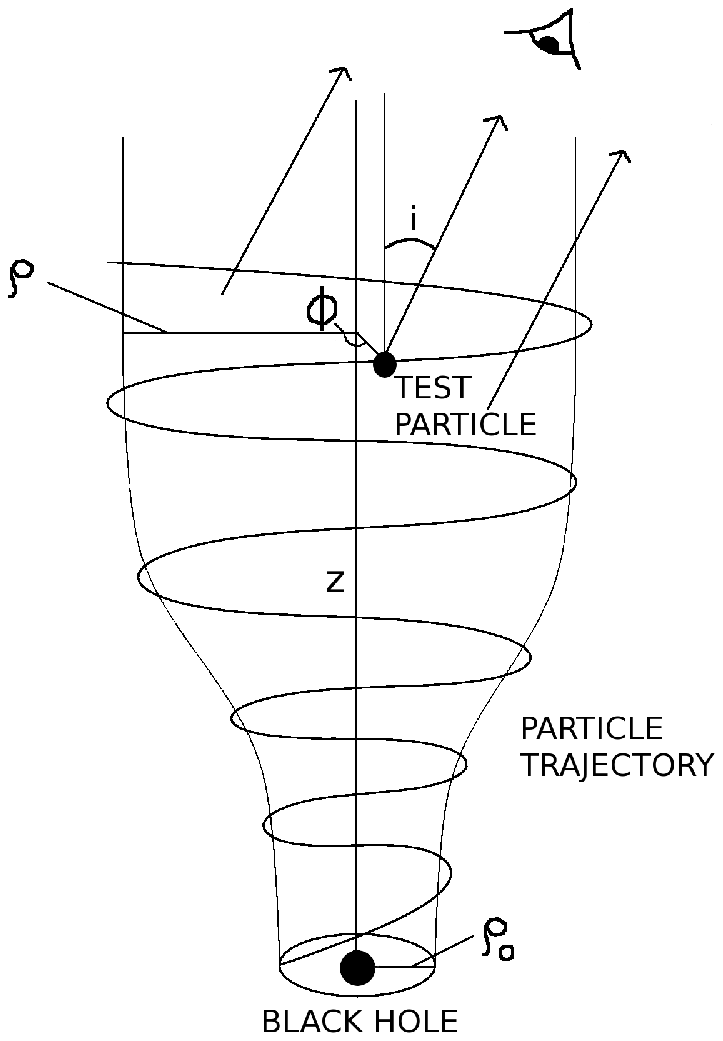}}
\centerline{\includegraphics[height=10.9cm,width=8cm]{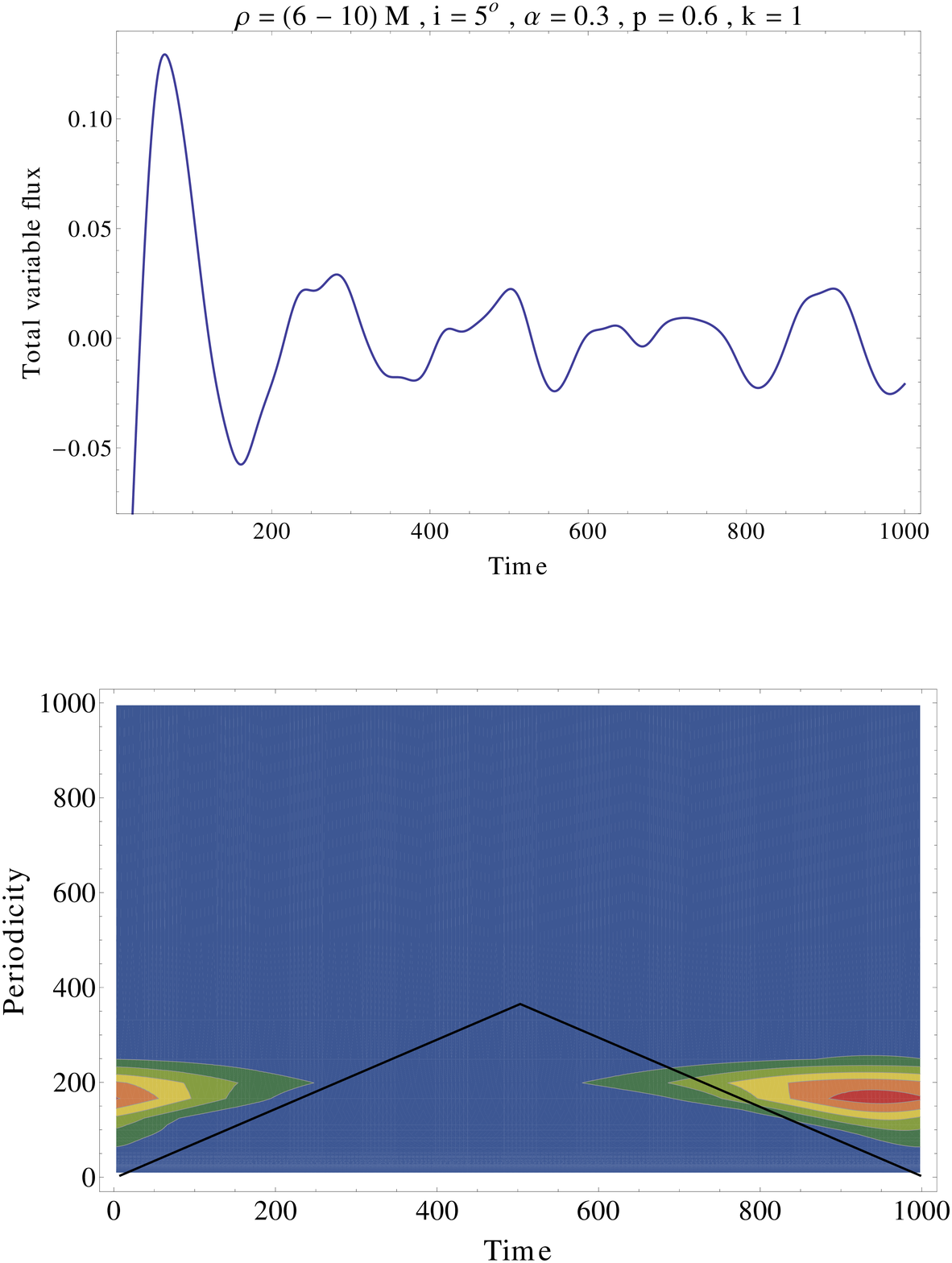}}
\caption{\small Top plot: helical trajectory of an emitting test particle in Schwarzschild geometry, constrained along rotating magnetic field lines with foot points on a Keplerian disk (at cylindrical radius $\rho_o$) and asymptotically bound by a cylinder of radius $\rho$ at large $z$. Middle plot: simulated combined light curve (total flux = $\sum^{i=5}_{i=1} \mathrm{flux}_i$, each $i$ corresponds to a $\rho_o$ between 6 M to 10 M) from emitting sources in helical motion with parameters $i=5^o$, $\alpha=0.3$, $p=0.6$ and $k=1$. Bottom plot: wavelet analysis of the total flux based time series.}
\label{fig2}
\end{figure}

Here, we present preliminary results based on extensions of the model proposed by Pech{\'a}{\v c}ek et al. (2005). We use models (cylindrical and funnel geometries) approximating the emergence of a jet, aimed at a unified view of variable emission from the inner region. The results are applicable to Seyferts and NLS1s with strong disk based emission as well as blazars with strong jet based emission. It can address variability induced features in LCs and periodograms and provide conditions for the presence and sustenance of an orbital motion (source) based QPO. With a Keplerian azimuthal rotational velocity $\Omega = \rho_o^{-\frac{3}{2}}$ where $\rho_o$ is the cylindrical distance to the footpoint from which a test particle on the disk begins its motion, the parametrization: $x = \rho_o \cos \phi$, $y = \rho_o \sin \phi$ is used for cylindrical geometry and $x = \rho_o (1+k (1-e^{-\alpha z})) \cos \phi$, $y = \rho_o (1+k (1-e^{-\alpha z})) \sin \phi$ is used for funnel geometry where $\phi$ is the azimuthal rotation angle, $k$ is an expansion factor and $\alpha$ is the rate of fall of the exponential term. In both these geometries, $z = p \phi$ where $p$ is the pitch. The QPO is present in all cases for a single component treatment and is diluted, broad and short lived for multiple components (Fig. \ref{fig2}). Beaming lasts for a short duration of 1 - 2 cycles at the start of the source motion and is more pronounced in the funnel model: (maximum flux of beamed portion)/(maximum of the Doppler component)$\sim$ 2. The later region is dominated by the Doppler component in both models. 
For a blazar with $M_8 = 1,  \rho_o = 6$ and small $z$, the observed timescale $T \sim 12.5 (1+z)$ hours where $z$ is the cosmological redshift. Allowing for a range of field geometries, it could range between less than a day to a few days, lasting over 2-3 cycles. For Kerr BHs, the horizon is smaller so that $T$ is shorter. For many Seyferts, $M_8 \sim 0.01-0.1$, so that $T$ can be an hour or less. These results confirm simulations indicating similar variability in BL Lacs reported for special relativistic plasma motion in magnetized jets (eg. Camenzind \& Krockenberger 1992). Detailed future studies (Mohan \& Mangalam; in preparation) will include the treatment of jet launching and a self-consistent treatment of the magnetic field, inclusion of effects due to a spatially distributed source and the calculation of synthetic PSDs which can be compared with observations to determine limiting time-scales for orbital motion where the timescale $T \sim 0.86 M_8 \rho_o^{\frac{3}{2}}$ hours.

\section{Time series analysis}

Time series analysis of AGN LCs using a combination of techniques reveals a wide range of properties such as variability characteristics and its correlation amongst different wavelengths, noise properties and the presence of a QPO and its evolution, thus aiding theoretical models. These techniques include: the periodogram, Lomb-Scargle periodogram (LSP), multi-harmonic analysis of variance (MHAoV) and wavelet analysis. The periodogram is the Fourier power spectrum of an evenly sampled, mean subtracted time series and is expressed in units of \\ (rms/variance)$^2$ (eg. Papadakis \& Lawrence 1993). It is useful in constraining the shape of the true expected distribution of Fourier power with frequencies, the power spectral density (PSD). The LSP \\ (Lomb 1976, Scargle 1982, Horne \& Baliunas 1986) is constructed using the algorithm suggested in Press \& Rybicki (1989) in order to achieve fast computational speeds. The multi-harmonic analysis of variance (MHAoV) is constructed using the algorithm suggested in Schwarzenberg-Czerny (1996). It has been applied to detect a QPO in the X-ray LC of the blazar 3C 454.3 (Lachowicz et al.\ 2009). The LSP and MHAoV can be used on both evenly and unevenly sampled data. In addition, the MHAoV can also detect non-sinusoidal signals. Wavelet analysis is used to both detect a QPO and determine the fraction of the total duration and hence the number of cycles that it is present for (e.g., Torrence \& Compo 1998; Oliver et al. 1998). Its development and usage, relationship with the Fourier transform, and various analyzing functions and their properties are reviewed in Farge (1992).
We first describe a set of numerical experiments which compare and bring out the advantages of each of these techniques.

\subsection{Numerical experiments}

Here, we compare the efficacy of each technique based on its robustness in detection of a periodicity and its complementary role in determination of the dataset properties. We test synthetic LCs with varying periodic signals: sinusoidal, $S$ or non-sinusoidal, $\bar{S}$; its sampling: even, $E$ or uneven $U$; its life-time as a fraction $f$ of the observation duration; the number of cycles it lasts during the observation $n_C=(t_P$/$P$) where $t_P$ is the duration of existence of the periodicity $P$; number of signals in the light curve, $n_S$ and the effects of introduction of noise. Table \ref{tab1} presents the results. The power peak $P_1$ in all experiments for the MHAoV and LSP techniques are consistently equal to or greater than the periodogram based peak. There is a power loss to other frequencies including harmonics of the primary frequency (spectral leakage) due to the finite length of the LC, the convolution of the LC with a rectangular window of duration equal to its length and sampling in the neighbourhood of $P$. This loss is least in the MHAoV technique in a majority of the experiments. A criteria for identifying a true periodicity is suggested to be $N_\sigma \geq 2$ for the periodogram and $N_\sigma \geq 4$ for the LSP and MHAoV. In $S, \ E$ and $\bar{S}, \ U$ datasets with noise, these criteria are met making the detection statistically significant. The experiments show that the techniques add up in a complementary manner to the finally available information, which cannot be obtained using any single technique. Noise is treated in the same manner for all techniques in the MC significance testing. Hence, if multiple techniques indicate a strong significance, the likelihood of existence of a QPO is high. In the experiments, the significance of false alarms are much smaller compared to 0.95 which is the criteria used to detect the periodicity $P$. The periodogram is used as the primary detection and noise characterization technique with a well developed analytic significance test. Significance testing for the LSP, MHAoV and wavelet analysis is carried out using Monte-Carlo simulations (see next section).

\subsection{Periodicity search strategy}
The below algorithm can be followed to detect QPOs and test for significance using Monte-Carlo (MC) simulations using the developed suite of analysis techniques:
\begin{enumerate}
\item Construct the periodogram/LSP and fit with an appropriate model (power law, power law with Lorentzian QPO, knee, break) using the Akaike information criteria and relative likelihood to determine the best fit model (see Mohan et al. 2013 in this volume). If a QPO is detected, determine its analytic significance. 
\item Use wavelet analysis to determine the phase during which the QPO exists and its duty cycle. 
\item Construct the MHAoV to confirm the detected QPO.
\item MC simulations based significance testing: use the Timmer \& Konig (1995) algorithm with the periodogram best fit model parameters and generate a large number (N $>$ 1000) of LCs. These are analyzed using all techniques. The number of instances in which the original power exceeds that obtained from random simulations divided by $N$ is reported as a significance value. 
\item Based on numerical experiments results, we suggest the criteria for detection of periodicity to be a significance $>$ 0.99.
\end{enumerate}

\subsection{Application, results and discussion}
\begin{figure}
\centerline{\includegraphics[height=9cm,width=20cm]{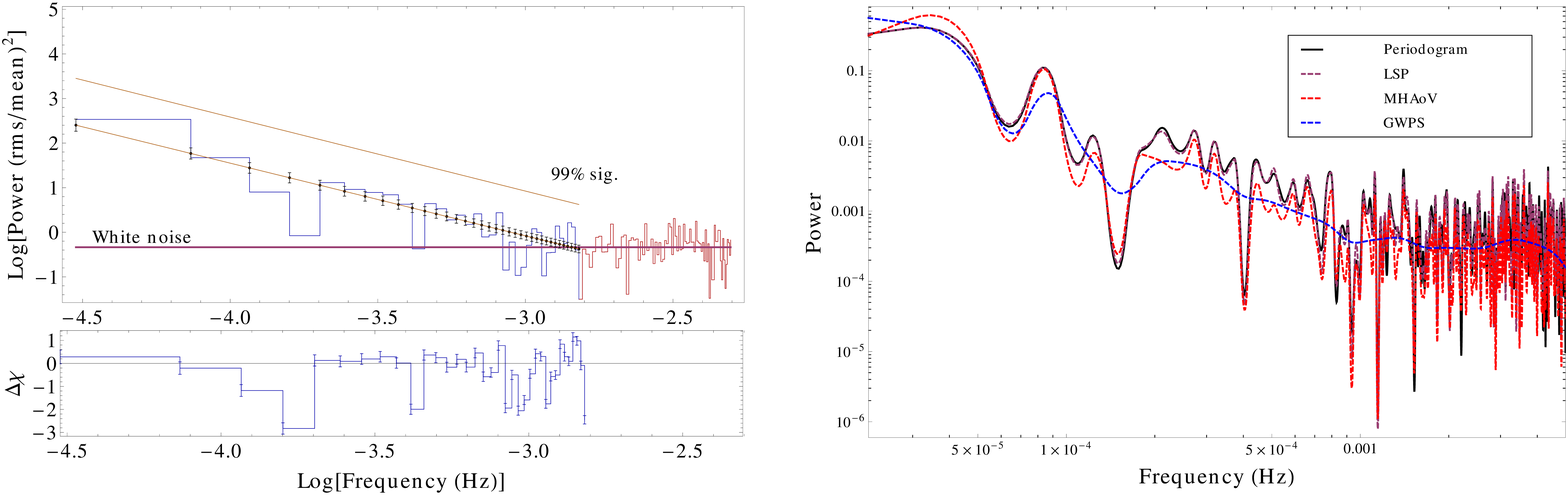}}
\caption{Results for S5 0716+714. Left plot: periodogram with power law fit. Right plot: combined plot showing smoothed periodogram, LSP, MHAoV and GWPS results. No significant periodic features are detected.}
\label{fig3}
\end{figure}

The search strategy is applied to a XMM Newton X-ray LC from S5 0716+714 (Obs. id. 0502271401) of duration $\sim$ 47000 s. The periodogram is best fit by a power law model (AIC = 63.8) in the red noise region with a slope of -1.63 $\pm$ 0.03. The closest model is the break frequency model (AIC = 69.1), the difference of 5.3 being large so as to rule it out. The results of the periodogram, LSP, MHAoV and wavelet analysis are plotted in Fig. \ref{fig3}. This dataset does not indicate any statistically significant results.

\subsection{Verification of time series techniques used}
In order to verify that the LSP and wavelet analysis give accurate results for real data sets, analytic expressions are computed for a standard cosine function of known periodicity. The functional form is then compared with numerical simulations.\\ 

{\em Lomb-Scargle periodogram}\\

Choosing $\disp{f(n)=\cos\frac{2 \pi n \Delta t}{N}}$ as the time series of length 2 N, along with $\Delta t = $ 1 s, the mean, $\overline{f} = 0$ and the variance, $\disp{\sigma^2 = \frac{2 N}{(2 N-1)}}$. The components of the LSP and $\tau$ are:
\begin{eqnarray} 
A (\omega) &=& \frac{(1-e^{i 2 \omega N})(\cos \frac{2 \pi}{N} e^{-i \omega}-1)}{e^{-2 i \omega}+1-2 e^{-i \omega} \cos \frac{2 \pi}{N}}\\ \nonumber
B (\omega) &=& \frac{(1-e^{-i 2 \omega N})(\cos \frac{2 \pi}{N} e^{i \omega}-1)}{e^{2 i \omega}+1-2 e^{i \omega} \cos \frac{2 \pi}{N}} \\ \nonumber
\end{eqnarray}
\begin{eqnarray} 
C (\omega) &=&\frac{1-e^{i 4 N \omega}}{e^{-i 2 \omega}-1} \\ \nonumber
D (\omega) &=&\frac{1-e^{-i 4 N \omega}}{e^{i 2 \omega}-1} \\ \nonumber
\tau (\omega) &=& \frac{1}{2\omega} \tan^{-1}\left[-i\frac{C(\omega)-D(\omega)}{C(\omega)+D(\omega)}\right]\\ \nonumber
\end{eqnarray}
\begin{eqnarray} 
\sum^{2 N}_{n=1} f(n) \cos \omega(n-\tau) = \frac{1}{2} (A (\omega) e^{-i \omega \tau}+B (\omega) e^{i \omega \tau}) \\ \nonumber
\sum^{2 N}_{n=1} f(n) \sin \omega(n-\tau) = \frac{1}{2 i} (A (\omega) e^{-i \omega \tau}-B (\omega) e^{i \omega \tau}) \\ \nonumber
\end{eqnarray}
\begin{eqnarray} 
\sum^{2 N}_{n=1} \cos^2 \omega(n-\tau) &=& N +\frac{C (\omega)}{4} e^{-2 i \omega \tau}\\ \nonumber
&+&\frac{D (\omega)}{4} e^{2 i \omega \tau} \\ \nonumber
\sum^{2 N}_{n=1} \sin^2 \omega(n-\tau) &=& N-\frac{C (\omega)}{4} e^{-2 i \omega \tau}-\frac{D (\omega)}{4} e^{2 i \omega \tau}.
\end{eqnarray}
Using the above relations, the power spectrum $P_N(\omega)$ (eg. Horne \& Baliunas 1986) can be constructed as a function of a suitable range of sampling frequencies, $\omega$.\\ 

{\em Wavelet analysis}\\

Choosing $x_n= \cos \omega_{k_0} t_n$ as the time series of length $N$ with $t_n=n \Delta t$, $k_0$ and $\disp{\omega_{k_0}=\frac{2 \pi k_0}{N \Delta t}}$, the discrete Fourier transform is $\disp{\hat{x}_k=\frac{\delta_{k,k_0}}{2}}$. The wavelet transform (using a Morlet wavelet sampling function), wavelet power spectrum (WPS) and the global wavelet power spectrum (GWPS) are
\begin{equation}
 W_n(s)=\frac{1}{2} \left(\frac{2\pi s}{\Delta t}\right)^{\frac{1}{2}} (\pi)^{-\frac{1}{4}} e^{-\frac{(s\omega_{k_0}-\omega_0)^2}{2}} e^{i\frac{2\pi k n}{N}}
\end{equation}
\begin{eqnarray}
WPS_x(s,n) &=& W_n(s) \cdot W^*_n(s) \\ \nonumber
&=& \frac{1}{4} \frac{2\pi s}{\Delta t} (\pi)^{-\frac{1}{2}} e^{-(s\omega_{k_0}-\omega_0)^2},
\end{eqnarray}
\begin{eqnarray}
GWPS_X(s) &=& \frac{1}{N} \sum^{N-1}_{n=0} W_n(s) \cdot W^*_n(s) \\ \nonumber
 &=& \frac{1}{4} \frac{2\pi s}{\Delta t} (\pi)^{-\frac{1}{2}} e^{-(s\frac{2\pi k_0}{N}-\omega_0)^2}.
\end{eqnarray}

The comparison between analytic and numerical results was carried out and are found to be in agreement.

\section{Conclusions}

Variability in AGN LCs is ubiquitous, spanning a large range of wavelengths. We have addressed variability due to orbital features in the inner region in the vicinity of the SMBH. Effects considered in models include: light bending, aberration, time delay, eclipsing due to disk structure, gravitational and \\ Doppler redshifts. We aim to develop detailed models which include contribution to the variability from the disk-jet region. Constraints on the emission region, characteristic timescales, the mass of the SMBH and their relationship were presented. Time series analysis of AGN LCs can be used to extract information on the emission mechanism, presence of break and knee frequencies or a QPO, its duration and phase and other aspects. Each technique was briefly discussed and numerical experiments (\S 4.1 and Table \ref{tab1}) making a comparative study were useful in preparing a search strategy for detection of QPOs (\S 4.2). Analytic expressions for the LSP and wavelet analysis were derived which can be shown to be in close agreement to numerical results. We aim to predict the theoretical properties of variable emission from analysis of observational results.

\bibliography{}

\onecolumn

\begin{table}
\resizebox{16.5cm}{6.5cm}{
\begin{tabular}{|l|l|l|l|l|l|l|}
\hline 
Expt. No.                                      & Technique   & Periodicity & MC         & Power & Width at         & $N_\sigma$ \\  
\&                                             &             & (s)         & sims.      & peak  & base of          &           \\
factors                                        &             &             & significance  & $P_1$ & $P_1$ (Hz)       &           \\  \hline
                                               &             &             &            &       &                  & \\
1) $\bar{S}$ (saw-tooth waveform), $E$, $n_S = 1$& Periodogram & 20          & -          & 0.62  & 0.02           & - \\
total duration of 100 s ($n_C = 5$)          & LSP         & 20          & -          & 0.62  & 0.02             & - \\
$f = 1$                                      & MHAoV       & 20          & -          & 0.80  & 0.03             & - \\
                                               & Wavelet     & 20          & -          & 0.41  & 0.06             & - \\ \hline
                                               &             &             &            &       &                  & \\
2) $S$ (sinusoidal waveform), $E$, $n_S = 1$    & Periodogram & 20          & 0.995      & 0.42  & 0.04           & 4.67\\
with noise                                     & LSP         & 20          & 0.995      & 0.42  & 0.04             & 4.67\\
total duration of 100 s ($n_C = 5$)            & MHAoV       & 20          & 0.996      & 0.62  & 0.04             & 5.88\\
$f = 1$                                       & Wavelet     & 20          & 0.996      & 0.25  & 0.06             & 4.89\\ \hline
                                               &             &             &            &       &                  & \\
3) $S$ (sinusoidal waveform), $U$, $n_S = 1$  & Periodogram & 20          & -          & 0.41  & 0.04             & - \\
total duration of 100 s, 20 points removed.    & LSP         & 20          & -          & 0.74  & 0.03             & - \\
                                               & MHAoV       & 20          & -          & 0.64  & 0.03             & - \\ \hline 
                                               &             &             &            &       &                  & \\
4) $\bar{S}$ (saw-tooth waveform), $U$, $n_S = 1$& Periodogram & 20        & 0.963      & 0.24  & 0.03             & 2.41\\
with noise. total duration of 100 s, 40 points & LSP         & 20          & 0.994      & 0.28  & 0.02             & 4.02\\
removed.                                       & MHAoV       & 20          & 0.995      & 0.58  & 0.02             & 6.36\\ \hline
                                               &             &             &            &       &                  & \\
5) $S$ (sinusoidal waveform), $E$, $n_S = 1$   & Periodogram & 20          & -          & 0.14  & 0.02             & - \\
total duration of 200 s ($f=0.5$, $n_C=5$)     & LSP         & 20          & -          & 0.14  & 0.02             & - \\
                                               & MHAoV       & 20          & -          & 0.19  & 0.02             & - \\
                                               & Wavelet     & 20          & -          & 0.09  & 0.06             & - \\ \hline
                                               &             &             &            &       &                  & \\
6) $S$ (sinusoidal waveform), $E$, $n_S = 2$   &Periodogram & 50$^2$, 100$^1$ &-&0.16$^2$, 0.36$^1$ & 0.01$^2$, 0.014$^1$ & - \\
total duration of 500 s                        & LSP & 50$^2$, 100$^1$ & - &0.16$^2$, 0.36$^1$ & 0.01$^2$, 0.014$^1$      & - \\
$f=0.4$, $n_C=4$ for 50 s periodicity       & MHAoV & 50$^2$, 100$^1$     & -       &0.15$^2$, 0.55$^1$ & 0.01$^2$, 0.012$^1$ & - \\
$f=0.6$, $n_C=3$ for 100 s periodicity   & Wavelet & 50$^2$, 100$^1$     & -       &0.17$^2$, 0.27$^1$ & 0.024$^2$, 0.014$^1$ & - \\
\hline
\end{tabular}}
\caption{The table shows a comparison of different techniques applied to a variety of waveforms. These results form the basis for the search strategy in \S 4.2. $^1$ Results for $P_1$;  $^2$ Results for $P_2$. $\disp{N_\sigma = \mathrm{\frac{peak \ power-mean}{standard \ deviation}}}$}
\label{tab1}
\end{table}



\end{document}